\documentclass[conference]{IEEEtran}
\ifCLASSINFOpdf
\else
\fi
\usepackage{amsmath}
\usepackage[nospace,noadjust]{cite}
\usepackage{amsfonts}
\usepackage{caption}
\usepackage[table]{xcolor} 
\usepackage{multirow}
\usepackage{booktabs}
\usepackage{pgfplots}
\usepackage{smartdiagram}
\usepackage{metalogo}
\usepackage{balance}
\usepackage[caption=false]{subfig}
\usepgfplotslibrary{units}
\usepgfplotslibrary{groupplots}
\usetikzlibrary{patterns}
\usepackage{tikz}
\usepackage{pgfplots}
\usetikzlibrary{colorbrewer}
\usepgfplotslibrary{colormaps}
\usetikzlibrary{calc,positioning,shapes,decorations.pathreplacing,snakes}
\usepackage{graphicx}
\DeclareCaptionLabelSeparator{periodspace}{.\quad}
\usepackage[nolist]{acronym}

\DeclareMathOperator*{\argmin}{arg\,min}

\pgfplotsset{colormap/jet}

\usepackage{url}
\pgfplotsset{compat=1.14}

\begin{document}
%
\title{SAIFE: Unsupervised Wireless Spectrum Anomaly Detection with Interpretable Features}


\author{\IEEEauthorblockN{Sreeraj Rajendran\IEEEauthorrefmark{1}, Wannes Meert\IEEEauthorrefmark{2}, Vincent Lenders\IEEEauthorrefmark{3} and Sofie Pollin\IEEEauthorrefmark{1}}
\\
\IEEEauthorblockA{
Email: \{sreeraj.rajendran, sofie.pollin\}@esat.kuleuven.be\\
wannes.meert@cs.kuleuven.be, vincent.lenders@armasuisse.ch.\\
\IEEEauthorrefmark{1}Department ESAT, KU Leuven, Belgium,
\IEEEauthorrefmark{2}Department of Computer Science, KU Leuven, Belgium\\
\IEEEauthorrefmark{3}armasuisse, Thun, Switzerland}}


%


\maketitle

\begin{abstract}
Detecting anomalous behavior in wireless spectrum is a demanding task due to the sheer complexity of the electromagnetic spectrum use. Wireless spectrum anomalies can take a wide range of forms from the presence of an unwanted signal in a licensed band to the absence of an expected signal, which makes manual labeling of anomalies difficult and suboptimal. We present, \ac{saif}, an \ac{aae} based anomaly detector for wireless spectrum anomaly detection using \ac{psd} data which achieves good anomaly detection and localization in an unsupervised setting. In addition, we investigate the model's capabilities to learn interpretable features such as signal bandwidth, class and center frequency in a semi-supervised fashion. Along with anomaly detection the model exhibits promising results for lossy \ac{psd} data compression up to 120X and semi-supervised signal classification accuracy close to 100\% on three datasets just using 20\% labeled samples. Finally the model is tested on data from one of the distributed Electrosense sensors over a long term of 500 hours showing its anomaly detection capabilities.

\end{abstract}


%
\IEEEpeerreviewmaketitle

\begin{acronym}[HBCI]
%
%
%
%
%

\acro{3gpp}[3GPP]{3\textsuperscript{rd} Generation Partnership Program}
\acro{cnn}[CNN]{Convolutional Neural Network}
\acro{fbmc}[FBMC]{Filter Bank Multicarrier}
\acro{phy}[PHY]{physical layer}
\acro{pu}[PU]{Primary User}
\acro{rat}[RAT]{Radio Access Technology}
\acro{rfnoc}[RFNoC]{RF Network on Chip}
\acro{sdr}[SDR]{Software Defined Radio}
\acro{su}[SU]{Secondary User}
\acro{toa}[TOA]{Time of Arrival}
\acro{tdoa}[TDOA]{Time Difference of Arrival}
\acro{usrp}[USRP]{Universal Software Radio Peripheral}
\acro{amc}[AMC]{Automatic Modulation Classification}
\acro{lstm}[LSTM]{Long Short Term Memory}
\acro{soa}[SoA]{state-of-the-art}
\acro{fft}[FFT]{Fast Fourier Transform}
\acro{wsn}[WSN]{Wireless Sensor Networks}
\acro{iq}[IQ]{In-phase and quadrature phase}
\acro{snr}[SNR]{signal-to-noise ratio}
\acro{sps}[sps]{samples/symbol}
\acro{awgn}[AWGN]{Additive White Gaussian Noise}
\acro{ofdm}[OFDM]{Orthogonal Frequency Division Multiplexing}
\acro{los}[LOS]{line of sight}
\acro{psd}[PSD]{Power Spectral Density}
\acro{svm}[SVM]{Support Vector Machines}
\acro{aae}[AAE]{Adversarial Autoencoder}
\acro{vae}[VAE]{Variational Autoencoder}
\acro{saif}[SAIFE]{Spectrum Anomaly Detector with Interpretable FEatures}
\acro{roc}[ROC]{Receiver operating characteristic}
\acro{auc}[AUC]{Area Under Curve}
\acro{dsa}[DSA]{Dynamic Spectrum Access}
\acro{hmm}[HMM]{Hidden Markov Models}
\end{acronym}
\section{Introduction}
\label{intro}

The new generation of wireless technologies 
is promising improved 
throughput, latency and reliability enabling the creation of novel applications. 
The fifth generation wireless deployments will be very heterogeneous ranging from millimeter wave communications to massive MIMO and LoRa/Sigfox deployments for \ac{los}, medium and long range communication systems respectively. Such dense and heterogeneous deployment makes the enforcement and management of the wireless spectrum usage difficult. In addition, manual spectrum management is inefficient and can only deal with a limited number of anomalies and measurement locations. Complex spectrum regulations across frequency bands in various countries along with illegal interference worsen the problem. Automated spectrum monitoring solutions covering frequency, time and space dimensions are becoming more crucial than ever before.

Unlike other sensing contexts such as air quality, temperature or city traffic monitoring, wireless spectrum monitoring on a large scale raises many unique problems ranging from the data costs associated with the sheer volume of sensed spectrum information to sensor quality and data privacy issues. These wide ranging infrastructure problems were systematically analyzed and partially solved by the Electrosense\footnote{\url{https://electrosense.org/}}\cite{electrosense} platform. Electrosense is interdisciplinary and combines the power of crowdsourcing  with Big data to solve the wireless spectrum monitoring problem. The sensing devices are low cost \ac{sdr} dongles connected to embedded devices like a Raspberry Pi or high end \ac{sdr} devices connected through a personal computer. Through Electrosense, an Open Spectrum Data as a Service (OSDaaS) model was introduced to address the usability of the spectrum data for a wide range of stakeholders including wireless operators, spectrum enforcement agencies, military and generic users.

In addition to the sensor infrastructure problems that were tackled in Electrosense, various algorithmic challenges still need to be addressed to provide advanced spectrum utilization awareness. The central coup to achieve this vision is a wireless spectrum anomaly detector which can continuously monitor the spectrum and detect unexpected behavior. Furthermore, in addition to the detection of anomalies, it is important to understand the cause of an anomaly. This ranges from an unexpected transmission in the analyzed band that can be classified \cite{lstm_classif}, to absence of an expected signal. Wireless anomaly detection to some extent has been addressed in wireless sensor networks in the past \cite{anomalywsn1,anomalywsn2,anomdetect3}. These techniques make use of derived expert features from very low rate sensor data such as temperature and pressure instead of high volume radio physical layer data as is our interest. An anomaly detector for \ac{dsa} is presented in \cite{aldoanomaly}, where distributed power measurements via cooperative sensing are used for anomaly detection. The proposed detector is limited to authorized user anomaly detection only, for the specific case of \ac{dsa}. Similarly \cite{anomalycognitive} makes use of \ac{hmm} on spectral amplitude probabilities that can detect interference on the channel of interest again in the \ac{dsa} domain.

Recently in \cite{o2016recurrent}, the authors presented a recurrent anomaly detector based on predictive modeling of raw \ac{iq} data. The authors used a \ac{lstm} model for predicting the next 4 \ac{iq} samples from the past 32 samples and an anomaly is detected based on the prediction error. Even though this model works on raw physical layer data which requires no expert feature extraction, it is still not sufficiently automated and generic for practical anomaly detection. First, different copies of the same model need to be trained for different wireless bands such that the model is able to predict anomalies specific to the band of interest. For instance, an LTE signal in the FM broadcast band is definitely an anomaly thus preventing a single model to be trained on both bands. Second, the model does not extract any interpretable features to understand the cause of the anomaly. In \cite{tandiya2018deep}, the authors extend this prediction idea on spectrograms and test the model on some synthetic anomalies. A reconstruction based anomaly detector based on vanilla deep autoencoders is presented in \cite{feng2017anomaly}. This model lacks interpretable feature extraction properties like class labels which implies the need for training multiple copies of the same model on different bands.

In this paper we argue that, reconstruction based anomaly detection could be superior to prediction based techniques as prediction is a tougher problem than reconstruction in complex time series datasets. For instance, while digitally modulating signals the basic assumption is each constellation point is selected with equal probability to maximize the information transfer which makes the prediction of the future symbols difficult. On the other hand reconstruction of input data from compressed features is an easier problem if the model can efficiently capture the complex data distributions. 

We propose \ac{saif}, an \ac{aae} based model which fills the shortcomings of these \ac{soa} models. First, we show that a \emph{single model can be trained over multiple bands} in an unsupervised fashion avoiding the need for multiple copies of models on various bands. Second, the same model can be \emph{trained in a semi-supervised fashion for extracting interpretable features} such as signal bandwidth and position. Third, the reconstructed signal from the proposed model can be used for \emph{localizing anomalies} in the wireless spectrum. Furthermore we explore various other advantages of the model such as \emph{wireless data compression} and \emph{signal classification} which are significant contributions in contrast to the \ac{soa} models \cite{o2016recurrent,tandiya2018deep,feng2017anomaly}. 

The rest of the paper is organized as follows. The anomaly detection problem is clearly stated in Section~\ref{problem}. Section~\ref{model} explains the \ac{aae} model used for anomaly detection and the parameters used for training along with the dataset details. Section~\ref{anomdetect} details the performance results and discusses the advantages of the proposed model. Section~\ref{compressclassif} explores the signal compression and classification features of the model. Conclusions and future work are presented in Section~\ref{conclusion}.

\section{Problem Definition}
\label{problem}
\textbf{Given:} Let $X_S$ be the source time-series data, where $\mathbf{x} \in X_S $ could be either a complex \ac{iq} vector or a frequency-based \ac{psd} vector from any wireless frequency band. The dataset $X_S$ contains wireless signals that are assumed to be normal behavior. Thus the probability of anomalous behavior in this source dataset is assumed to be low. The superset $X_S = X_{S0} \cup X_{S1} ... \cup X_{Sn}$ contains signals from various frequency bands. 

\textbf{Goal:} A model that learns the source data distribution $p(X_S)$ and detects when a target vector's distribution deviates from the source data distribution. For each target vector $\mathbf{x} \in X_T$, $X_T$ being the test dataset, the model should infer whether the vector is normal~($H_0$) or anomalous ($H_a$), where $H_0$ and $H_a$ are hypothesis listed below.
\begin{itemize}
\item $H_0$: Sample data comes from $p(X_S)$
\item $H_a$: Sample data does not come from $p(X_S)$
\end{itemize}
A signal type of $\mathbf{x}\in X_{Sl}$ from frequency band $l$ occurring in a band $k$ where we are expecting $X_{Sk}$ is also an anomalous behavior which demands the model to capture class labels for fine grained anomaly detection.

\textbf{Assumptions:}
\begin{enumerate}
\item The probability of anomalous behavior in the source dataset is very low.
\item No explicit anomaly labeling is done on the source and target dataset.
\item No expert feature extraction is performed before feeding data to the model.
\end{enumerate}

\section{Models}
\label{model}
\begin{figure} [!t]
    \centering
  \subfloat[Vanilla autoencoder]{%
       \includegraphics[width=0.38\linewidth]{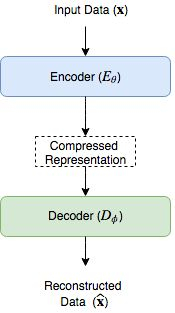}}
    \label{fig:autoencoder}\hfill
  \subfloat[Variational autoencoder]{%
        \includegraphics[width=0.40\linewidth]{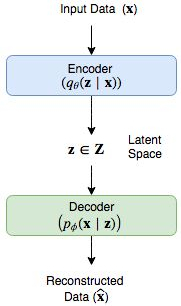}}
    \label{fig:vae}\hfill
   \caption{(a) Encoder decoder structure of an unsupervised vanilla autoencoder model, (b) Stochastic variant of the autoencoder where the internal representations are probability distributions in general 
}
  \label{fig:models} 
\end{figure}

We leverage the recent advances in generative modeling using neural networks which are trained through backpropagation directly from data \cite{vaekingma, makhzani2015adversarial, goodfellow2014generative, infogan}. The key insight of these previous work is to bring the higher dimensional input data to some lower dimensional latent space ($\mathbf{Z}$), whose prior distributions can be specified. This latent space which captures relevant features or settings can be then used to reconstruct the actual input data, ideally with minimal reconstruction loss. A basic introduction to some of the recent \ac{soa} generative models are covered in the following subsections. 

\subsection{Autoencoder and \ac{vae}}
A traditional autoencoder, as shown in Figure~\ref{fig:models}, is a neural network that consists of an encoder (E) and a decoder (D). The encoder and decoder are trained to reduce the reconstruction loss. This entire network basically performs a non-linear dimensionality reduction optimizing the encoder and decoder parameters ($\theta$ and $\phi$), the neural network weights, to achieve minimum reconstruction loss such as minimum squared error as given below
\begin{equation}
\theta,\phi= \argmin\limits_{{\theta,\phi}}||\mathbf{x} - \hat{\mathbf{x}}||^2
\end{equation}

A \ac{vae} \cite{vaekingma} also makes use of an its encoder-decoder structure. \ac{vae}s encode the input data vector to a vector $\mathbf{z}$ in the latent space $\mathbf{Z}$  whose priors can be imposed by using a Kullback-Leibler (KL) divergence penalty. \ac{vae} optimizes the network parameters $\theta$ and $\phi$ to minimize the following upper-bound on the negative log-likelihood of $\mathbf{x}$, where $p_{data}$ is the distribution of the data $\mathbf{x}$:
\begin{equation}
\begin{split}
\mathbb{E}_{\mathbf{x}\sim p_{data}}[-log~p_\phi(\mathbf{x})] &< \mathbb{E}_{\mathbf{x}\sim p_{data}}[\mathbb{E}_{\mathbf{z}\sim q_\theta(\mathbf{z}|\mathbf{x})}[-log(p_\phi(\mathbf{x}|\mathbf{z})]]\\ 
&+ \mathbb{E}_{\mathbf{x}\sim p_{data}}[KL(q_\theta(\mathbf{z}|\mathbf{x})||p_\phi(\mathbf{z}))]
\end{split}
\end{equation}

Thus \ac{vae} optimize the reconstruction loss (first term) similar to a standard autoencoder but adds regularization terms (second term: KL divergence or cross-entropy term) which helps it to learn a latent representation that is consistent with the defined prior $p_\phi(\mathbf{z})$.

\subsection{Adversarial autoencoder (\ac{aae})}
Adversarial autoencoders \cite{makhzani2015adversarial} make use of the recent advances in generative modeling \cite{goodfellow2014generative} to replace the KL divergence in \ac{vae}s with adversarial training that encourages the decoder to map the imposed prior to the data distribution. Thus \ac{aae} provides two major advantages over \ac{vae}: (i) the model ensures that the decoder will generate meaningful samples if we sample from any part of the prior space and (ii) as the aggregate posterior matches the prior distribution, variations of these distributions can be used for detecting unknown data inputs which is very useful for applications such as anomaly detection. In addition, \ac{aae} provides a flexible and robust architecture for semi-supervised learning and data visualization.

\subsection{\ac{saif} description}
\begin{figure}[t]
\centering
\includegraphics[width=\columnwidth]{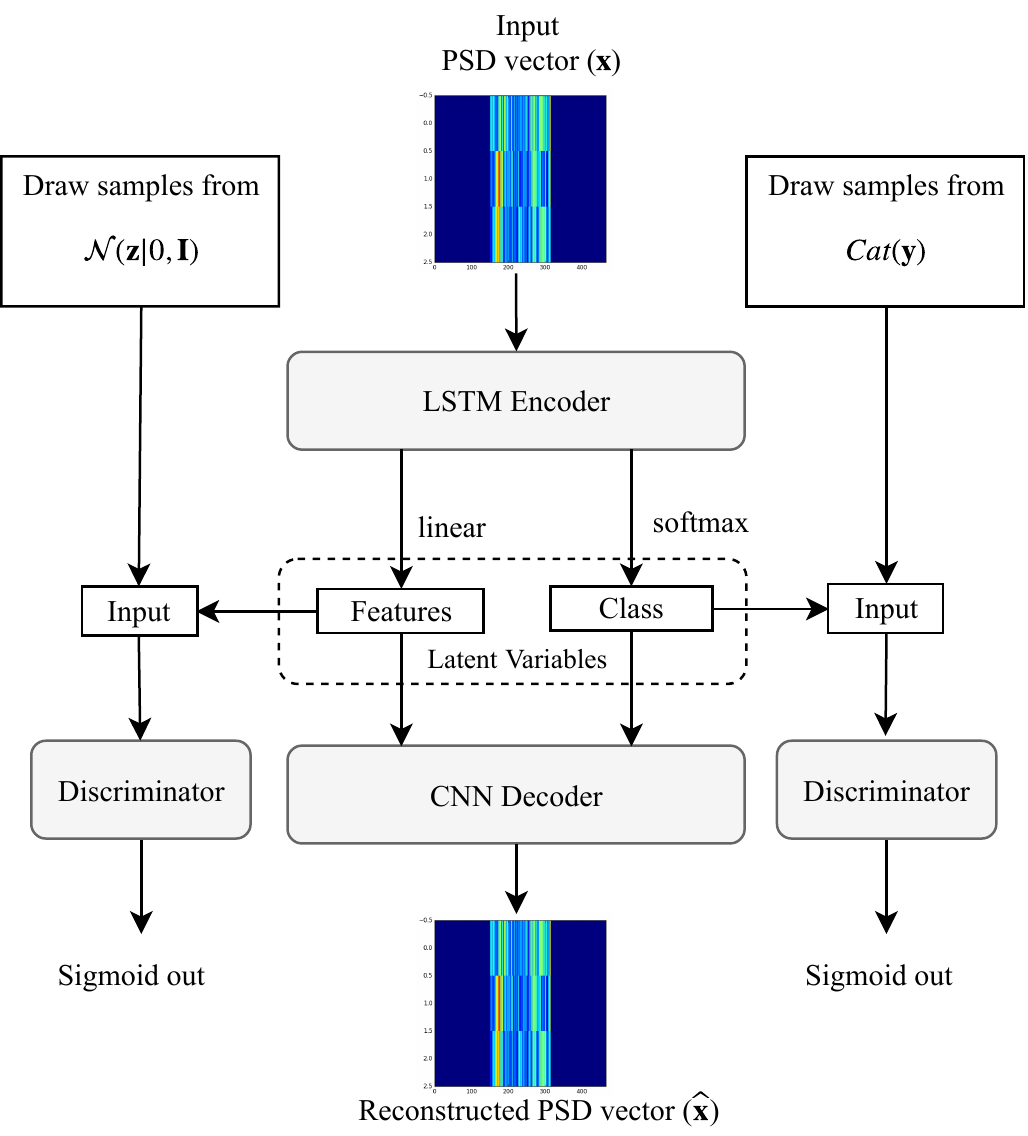}
\caption{Model architecture for anomaly detection.}
\label{fig:model_arch}
\end{figure}

We make use of a deep learning model based on \ac{aae} to enable all the requirements mentioned in the problem definition as shown in Figure~\ref{fig:model_arch}. An \ac{lstm} layer with 512 cells is used as the encoder for extracting interpretable features while a \ac{cnn} based decoder is employed for reconstructing the input data from the extracted features. The \ac{aae} architecture is trained in a semi-supervised fashion for making the features more interpretable while the reconstruction is fully  unsupervised. Two layer feed forward networks with 256 cells and relu activations are employed in both discriminators. The \ac{lstm} output is fed through a softmax layer for signal classification and a linear layer for extracting the latent features. 

The discriminators~($D_s$) are neural networks that evaluate the probability that the latent code $\mathbf{z}$ is from the prior distribution $p(\mathbf{z})$ that we are trying to impose rather than a sample from the output of the encoder~($E$) model. The discriminator receives $\mathbf{z}$ from both the encoder and the prior distribution and is trained to distinguish between them. The encoder is trained to confuse the discriminators into believing that the samples it generates are from the prior distribution. Thus the encoder is trained to reach the solution by optimizing both networks by playing a min-max adversarial game which is expressed in \cite{goodfellow2014generative} as 
\begin{equation}
\min_{E}\max_{D_s} \mathbb{E}_{\mathbf{z}\sim p(\mathbf{z})}[log(D_s(\mathbf{z}))]+\mathbb{E}_{\mathbf{x}\sim p_{data}}[log~(1-D_s(E(\mathbf{x})))] 
\end{equation}

Generative models try to model the underlying distributions of the input data, the latent variables, which are further used for data reconstruction. In \ac{saif}, the input \ac{psd} data is assumed to be generated by the latent \textit{Class} variable which comes from a Categorical distribution with number of categories $k=$\textit{number of frequency bands} and the continuous latent \textit{Features} from a Gaussian distribution of zero mean and unit variance; $p(\mathbf{y}) = Cat(\mathbf{y})$ and $p(\mathbf{z}) = \mathcal{N}(\mathbf{z}|0,\mathbf{I})$.

\subsection{Datasets}
\label{datasets}
\begin{figure}[!t]
\centering
\includegraphics[width=0.4\columnwidth]{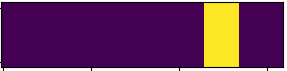}
\includegraphics[width=0.4\columnwidth]{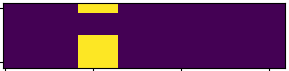}
\includegraphics[width=0.4\columnwidth]{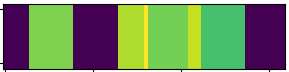}
\includegraphics[width=0.4\columnwidth]{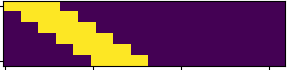}
\caption{Sample signals \textit{single-cont}, \textit{single-rshort}, \textit{mult-cont} and \textit{dethop} from synthetic signal dataset (time on y-axis and frequency on x-axis).}
\label{fig:syn_dataset}
\end{figure}

We use three spectrum datasets along with one synthetic anomaly set to evaluate the performance of the used model. A synthetic spectrum dataset is necessary to understand the performance of the model in a controlled environment. The synthetic dataset consists of four different signal types with signal parameters as reported in Table \ref{table_syn_dataset}. The signals being (i) \textit{single-cont}: single continuous signal with random bandwidth, \ac{snr} and center frequency, (ii) \textit{single-rshort}: pulsed signals in time with similar parameters as \textit{single-cont}, (iii) \textit{mult-cont}: multiple continuous signals with possible overlap and (iv) \textit{dethop}: random bandwidth and \ac{snr} signals with deterministic shifts/hops in frequency as depicted in Figure \ref{fig:syn_dataset}. Similarly, four synthetic signals (i) \textit{scont}: same as single-cont, (ii) \textit{randpulses}: random pulsed transmissions on the given band, (iii) \textit{wpulse}: pulsed wideband signals covering the entire frequency, (iv) \textit{oclass}: signals from other classes in synthetic dataset are used as anomalies.

\begin{table}[!t]
\begin{center}
\begin{tabular}{|l|l|}
	\hline
    Type     & single-cont, single-rshort, \\
             & mult-cont, dethop\\   								\hline
  	Input frame size &   6x64 \\
  	\hline
    SNR Range &  5dB to +20dB \\
  	\hline
   	Number of training samples &   6000 vectors\\
  	\hline
  	Number of test samples &  6000 vectors\\
    \hline
\end{tabular}
\end{center}
\caption{Synthetic signal dataset parameters.}
\label{table_syn_dataset}
\end{table}

\begin{table}[!t]
\begin{center}
\begin{tabular}{|l|l|}
	\hline
    Type     & scont, randpulses, wpulse, oclass\\   							\hline
  	Input frame size &   6x64 \\
  	\hline
    SNR Range &  -20dB to +20dB \\
  	\hline
   	Number of training samples &   6000 vectors\\
  	\hline
  	Number of test samples &  6000 vectors\\
    \hline
\end{tabular}
\end{center}
\caption{Synthetic anomaly dataset parameters.}
\label{table_syna_dataset}
\end{table}

\begin{figure}[t]
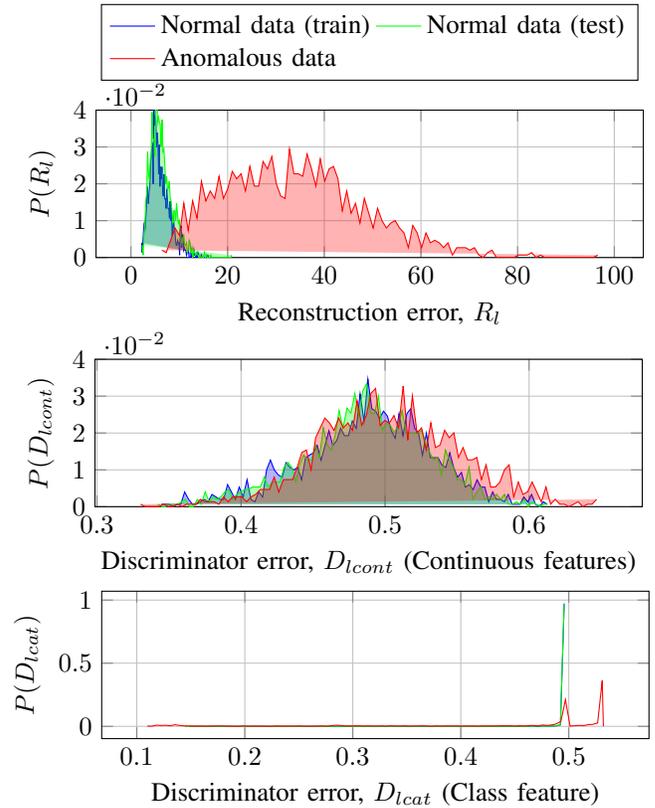

\centering
\input{figures/errdist.tex}
\input{figures/contdist.tex}
\begin{tikzpicture}
\begin{axis}[legend style={at={(0.5,1.57)},anchor=north},legend columns=2,width=\columnwidth,height=0.4\columnwidth, grid=both,xlabel={Discriminator error, $D_{lcat}$ (Class feature)},ylabel={$P(D_{lcat})$},grid style={line width=.1pt, draw=gray!10},major grid style={line width=.2pt,draw=gray!50},legend cell align={left},]
\addplot[color=blue, fill opacity=0.3] coordinates { 
( 0.148886412382 , 0.000664893617021 ) ( 0.152389812768 , 0.0 ) ( 0.155893213153 , 0.0 ) ( 0.159396613538 , 0.0 ) ( 0.162900013924 , 0.0 ) ( 0.166403414309 , 0.0 ) ( 0.169906814694 , 0.0 ) ( 0.17341021508 , 0.0 ) ( 0.176913615465 , 0.0 ) ( 0.180417015851 , 0.0 ) ( 0.183920416236 , 0.0 ) ( 0.187423816621 , 0.0 ) ( 0.190927217007 , 0.0 ) ( 0.194430617392 , 0.0 ) ( 0.197934017777 , 0.0 ) ( 0.201437418163 , 0.0 ) ( 0.204940818548 , 0.0 ) ( 0.208444218934 , 0.000664893617021 ) ( 0.211947619319 , 0.000664893617021 ) ( 0.215451019704 , 0.0 ) ( 0.21895442009 , 0.0 ) ( 0.222457820475 , 0.0 ) ( 0.22596122086 , 0.0 ) ( 0.229464621246 , 0.0 ) ( 0.232968021631 , 0.0 ) ( 0.236471422017 , 0.0 ) ( 0.239974822402 , 0.0 ) ( 0.243478222787 , 0.0 ) ( 0.246981623173 , 0.0 ) ( 0.250485023558 , 0.0 ) ( 0.253988423944 , 0.0 ) ( 0.257491824329 , 0.0 ) ( 0.260995224714 , 0.0 ) ( 0.2644986251 , 0.0 ) ( 0.268002025485 , 0.000664893617021 ) ( 0.27150542587 , 0.0 ) ( 0.275008826256 , 0.0 ) ( 0.278512226641 , 0.0 ) ( 0.282015627027 , 0.0 ) ( 0.285519027412 , 0.0 ) ( 0.289022427797 , 0.0 ) ( 0.292525828183 , 0.0 ) ( 0.296029228568 , 0.0 ) ( 0.299532628953 , 0.0 ) ( 0.303036029339 , 0.0 ) ( 0.306539429724 , 0.0 ) ( 0.31004283011 , 0.0 ) ( 0.313546230495 , 0.0 ) ( 0.31704963088 , 0.0 ) ( 0.320553031266 , 0.0 ) ( 0.324056431651 , 0.0 ) ( 0.327559832036 , 0.0 ) ( 0.331063232422 , 0.0 ) ( 0.334566632807 , 0.0 ) ( 0.338070033193 , 0.0 ) ( 0.341573433578 , 0.0 ) ( 0.345076833963 , 0.0 ) ( 0.348580234349 , 0.0 ) ( 0.352083634734 , 0.0 ) ( 0.35558703512 , 0.0 ) ( 0.359090435505 , 0.0 ) ( 0.36259383589 , 0.0 ) ( 0.366097236276 , 0.0 ) ( 0.369600636661 , 0.0 ) ( 0.373104037046 , 0.0 ) ( 0.376607437432 , 0.0 ) ( 0.380110837817 , 0.0 ) ( 0.383614238203 , 0.0 ) ( 0.387117638588 , 0.0 ) ( 0.390621038973 , 0.0 ) ( 0.394124439359 , 0.0 ) ( 0.397627839744 , 0.0 ) ( 0.401131240129 , 0.0 ) ( 0.404634640515 , 0.0 ) ( 0.4081380409 , 0.0 ) ( 0.411641441286 , 0.0 ) ( 0.415144841671 , 0.0 ) ( 0.418648242056 , 0.0 ) ( 0.422151642442 , 0.0 ) ( 0.425655042827 , 0.0 ) ( 0.429158443213 , 0.0 ) ( 0.432661843598 , 0.0 ) ( 0.436165243983 , 0.0 ) ( 0.439668644369 , 0.0 ) ( 0.443172044754 , 0.0 ) ( 0.446675445139 , 0.0 ) ( 0.450178845525 , 0.0 ) ( 0.45368224591 , 0.0 ) ( 0.457185646296 , 0.0 ) ( 0.460689046681 , 0.0 ) ( 0.464192447066 , 0.0 ) ( 0.467695847452 , 0.00132978723404 ) ( 0.471199247837 , 0.0 ) ( 0.474702648222 , 0.000664893617021 ) ( 0.478206048608 , 0.00265957446809 ) ( 0.481709448993 , 0.00265957446809 ) ( 0.485212849379 , 0.00265957446809 ) ( 0.488716249764 , 0.00332446808511 ) ( 0.492219650149 , 0.0113031914894 ) ( 0.495723050535 , 0.972739361702 )
};
\addplot[color=green, fill opacity=0.3] coordinates { 
( 0.144380018115 , 0.000668449197861 ) ( 0.147928482443 , 0.000668449197861 ) ( 0.151476946771 , 0.0 ) ( 0.155025411099 , 0.0 ) ( 0.158573875427 , 0.0 ) ( 0.162122339755 , 0.0 ) ( 0.165670804083 , 0.0 ) ( 0.169219268411 , 0.0 ) ( 0.172767732739 , 0.0 ) ( 0.176316197067 , 0.0 ) ( 0.179864661396 , 0.0 ) ( 0.183413125724 , 0.0 ) ( 0.186961590052 , 0.0 ) ( 0.19051005438 , 0.0 ) ( 0.194058518708 , 0.0 ) ( 0.197606983036 , 0.0 ) ( 0.201155447364 , 0.0 ) ( 0.204703911692 , 0.0 ) ( 0.20825237602 , 0.0 ) ( 0.211800840348 , 0.0 ) ( 0.215349304676 , 0.0 ) ( 0.218897769004 , 0.000668449197861 ) ( 0.222446233332 , 0.000668449197861 ) ( 0.22599469766 , 0.0 ) ( 0.229543161988 , 0.000668449197861 ) ( 0.233091626316 , 0.0 ) ( 0.236640090644 , 0.0 ) ( 0.240188554972 , 0.0 ) ( 0.2437370193 , 0.0 ) ( 0.247285483629 , 0.0 ) ( 0.250833947957 , 0.000668449197861 ) ( 0.254382412285 , 0.0 ) ( 0.257930876613 , 0.0 ) ( 0.261479340941 , 0.0 ) ( 0.265027805269 , 0.0 ) ( 0.268576269597 , 0.0 ) ( 0.272124733925 , 0.0 ) ( 0.275673198253 , 0.0 ) ( 0.279221662581 , 0.0 ) ( 0.282770126909 , 0.0 ) ( 0.286318591237 , 0.0 ) ( 0.289867055565 , 0.0 ) ( 0.293415519893 , 0.0 ) ( 0.296963984221 , 0.000668449197861 ) ( 0.300512448549 , 0.0 ) ( 0.304060912877 , 0.0 ) ( 0.307609377205 , 0.0 ) ( 0.311157841533 , 0.0 ) ( 0.314706305861 , 0.0 ) ( 0.31825477019 , 0.0 ) ( 0.321803234518 , 0.0 ) ( 0.325351698846 , 0.0 ) ( 0.328900163174 , 0.0 ) ( 0.332448627502 , 0.0 ) ( 0.33599709183 , 0.0 ) ( 0.339545556158 , 0.0 ) ( 0.343094020486 , 0.0 ) ( 0.346642484814 , 0.0 ) ( 0.350190949142 , 0.000668449197861 ) ( 0.35373941347 , 0.0 ) ( 0.357287877798 , 0.0 ) ( 0.360836342126 , 0.0 ) ( 0.364384806454 , 0.0 ) ( 0.367933270782 , 0.0 ) ( 0.37148173511 , 0.0 ) ( 0.375030199438 , 0.0 ) ( 0.378578663766 , 0.0 ) ( 0.382127128094 , 0.0 ) ( 0.385675592422 , 0.0 ) ( 0.389224056751 , 0.0 ) ( 0.392772521079 , 0.0 ) ( 0.396320985407 , 0.0 ) ( 0.399869449735 , 0.000668449197861 ) ( 0.403417914063 , 0.0 ) ( 0.406966378391 , 0.0 ) ( 0.410514842719 , 0.0 ) ( 0.414063307047 , 0.0 ) ( 0.417611771375 , 0.0 ) ( 0.421160235703 , 0.0 ) ( 0.424708700031 , 0.0 ) ( 0.428257164359 , 0.0 ) ( 0.431805628687 , 0.000668449197861 ) ( 0.435354093015 , 0.0 ) ( 0.438902557343 , 0.000668449197861 ) ( 0.442451021671 , 0.000668449197861 ) ( 0.445999485999 , 0.0 ) ( 0.449547950327 , 0.0 ) ( 0.453096414655 , 0.000668449197861 ) ( 0.456644878983 , 0.0 ) ( 0.460193343312 , 0.0 ) ( 0.46374180764 , 0.0 ) ( 0.467290271968 , 0.000668449197861 ) ( 0.470838736296 , 0.00267379679144 ) ( 0.474387200624 , 0.000668449197861 ) ( 0.477935664952 , 0.0 ) ( 0.48148412928 , 0.00133689839572 ) ( 0.485032593608 , 0.00267379679144 ) ( 0.488581057936 , 0.00601604278075 ) ( 0.492129522264 , 0.0106951871658 ) ( 0.495677986592 , 0.966577540107 )
};
\addplot[color=red, fill opacity=0.3] coordinates { 
( 0.109713174403 , 0.00200534759358 ) ( 0.113967952654 , 0.00133689839572 ) ( 0.118222730905 , 0.00935828877005 ) ( 0.122477509156 , 0.00668449197861 ) ( 0.126732287407 , 0.00868983957219 ) ( 0.130987065658 , 0.00467914438503 ) ( 0.135241843909 , 0.0127005347594 ) ( 0.13949662216 , 0.00735294117647 ) ( 0.143751400411 , 0.00401069518717 ) ( 0.148006178662 , 0.00267379679144 ) ( 0.152260956913 , 0.00133689839572 ) ( 0.156515735164 , 0.00200534759358 ) ( 0.160770513415 , 0.00200534759358 ) ( 0.165025291666 , 0.000668449197861 ) ( 0.169280069917 , 0.00200534759358 ) ( 0.173534848168 , 0.00267379679144 ) ( 0.17778962642 , 0.0033422459893 ) ( 0.182044404671 , 0.00200534759358 ) ( 0.186299182922 , 0.00133689839572 ) ( 0.190553961173 , 0.000668449197861 ) ( 0.194808739424 , 0.0 ) ( 0.199063517675 , 0.0033422459893 ) ( 0.203318295926 , 0.000668449197861 ) ( 0.207573074177 , 0.00133689839572 ) ( 0.211827852428 , 0.000668449197861 ) ( 0.216082630679 , 0.00200534759358 ) ( 0.22033740893 , 0.000668449197861 ) ( 0.224592187181 , 0.0 ) ( 0.228846965432 , 0.00267379679144 ) ( 0.233101743683 , 0.00200534759358 ) ( 0.237356521934 , 0.00133689839572 ) ( 0.241611300185 , 0.00200534759358 ) ( 0.245866078436 , 0.00133689839572 ) ( 0.250120856687 , 0.0033422459893 ) ( 0.254375634938 , 0.000668449197861 ) ( 0.25863041319 , 0.00133689839572 ) ( 0.262885191441 , 0.00133689839572 ) ( 0.267139969692 , 0.00133689839572 ) ( 0.271394747943 , 0.0033422459893 ) ( 0.275649526194 , 0.000668449197861 ) ( 0.279904304445 , 0.00467914438503 ) ( 0.284159082696 , 0.00868983957219 ) ( 0.288413860947 , 0.00735294117647 ) ( 0.292668639198 , 0.0033422459893 ) ( 0.296923417449 , 0.0033422459893 ) ( 0.3011781957 , 0.00401069518717 ) ( 0.305432973951 , 0.00401069518717 ) ( 0.309687752202 , 0.00200534759358 ) ( 0.313942530453 , 0.00401069518717 ) ( 0.318197308704 , 0.00200534759358 ) ( 0.322452086955 , 0.00200534759358 ) ( 0.326706865206 , 0.00401069518717 ) ( 0.330961643457 , 0.00133689839572 ) ( 0.335216421708 , 0.00467914438503 ) ( 0.33947119996 , 0.00267379679144 ) ( 0.343725978211 , 0.00200534759358 ) ( 0.347980756462 , 0.00467914438503 ) ( 0.352235534713 , 0.00133689839572 ) ( 0.356490312964 , 0.00267379679144 ) ( 0.360745091215 , 0.00267379679144 ) ( 0.364999869466 , 0.00467914438503 ) ( 0.369254647717 , 0.00133689839572 ) ( 0.373509425968 , 0.000668449197861 ) ( 0.377764204219 , 0.0033422459893 ) ( 0.38201898247 , 0.00133689839572 ) ( 0.386273760721 , 0.0033422459893 ) ( 0.390528538972 , 0.00267379679144 ) ( 0.394783317223 , 0.0033422459893 ) ( 0.399038095474 , 0.00200534759358 ) ( 0.403292873725 , 0.00200534759358 ) ( 0.407547651976 , 0.00267379679144 ) ( 0.411802430227 , 0.00267379679144 ) ( 0.416057208478 , 0.00133689839572 ) ( 0.42031198673 , 0.0033422459893 ) ( 0.424566764981 , 0.0033422459893 ) ( 0.428821543232 , 0.0033422459893 ) ( 0.433076321483 , 0.00267379679144 ) ( 0.437331099734 , 0.00467914438503 ) ( 0.441585877985 , 0.00467914438503 ) ( 0.445840656236 , 0.00467914438503 ) ( 0.450095434487 , 0.00200534759358 ) ( 0.454350212738 , 0.00534759358289 ) ( 0.458604990989 , 0.00200534759358 ) ( 0.46285976924 , 0.00601604278075 ) ( 0.467114547491 , 0.00735294117647 ) ( 0.471369325742 , 0.0033422459893 ) ( 0.475624103993 , 0.0106951871658 ) ( 0.479878882244 , 0.00735294117647 ) ( 0.484133660495 , 0.0106951871658 ) ( 0.488388438746 , 0.019385026738 ) ( 0.492643216997 , 0.0394385026738 ) ( 0.496897995248 , 0.210561497326 ) ( 0.501152773499 , 0.00200534759358 ) ( 0.505407551751 , 0.00467914438503 ) ( 0.509662330002 , 0.00735294117647 ) ( 0.513917108253 , 0.00601604278075 ) ( 0.518171886504 , 0.0133689839572 ) ( 0.522426664755 , 0.0160427807487 ) ( 0.526681443006 , 0.0267379679144 ) ( 0.530936221257 , 0.362299465241 ) (0.532, 0.0)
};
\end{axis}
\end{tikzpicture}
\caption{Probability density functions of reconstruction error, continuous discriminator error and categorical discriminator error for \textit{dethop} signal and \textit{dethop} signal with \textit{scont} anomaly.}
\label{fig:err_dist}
\end{figure}

In addition to the synthetic dataset we validate using two real wireless datasets. The first is a \ac{sdr} dataset collected using a HackRF SDR from two different cities in Belgium covering frequencies from 10~MHz to 3~GHz. HackRF with its firmware sweep mode can scan the spectrum at up to 8 GHz per second, which allows scanning of 0-6~GHz under a second. Twelve frequency bands are selected from these spectrum scans, continuous in time covering various audio and video broadcast, GSM and LTE bands with a frequency resolution of 100~KHz whose frequency ranges are listed in Table~\ref{table_dataset_bands}. The second dataset consists of \ac{psd} sensor data from multiple Electrosense sensors deployed all over Europe retrieved through the open API\footnote{\url{https://electrosense.org/open-api-spec.html}} with 7 selected frequency bands as listed in Table~\ref{table_dataset_bands}.

\begin{table}[htb]
\begin{center}
\begin{tabular}{|l|l|}
	\hline
    Dataset and bands    & Frequencies (MHz)\\   							\hline
  	SDR dataset, 0-11 &   80-107, 109-115.5, 117-140, 166-172,\\
    & 196-208, 212.5-217.5, 220-227.5, 388-396,\\ 
    & 422-427, 640-660, 790-800, 920-960.\\
  	\hline
    Electrosense dataset, 0-6 &  86-108, 192-197, 790-801,\\ 
    & 801-810, 811-821, 933-935, 955-960. \\
    \hline
\end{tabular}
\end{center}
\caption{SDR and Electrosense dataset frequency bands.}
\label{table_dataset_bands}
\end{table}

\subsection{Model training}
All the datasets mentioned in the previous section are split into two subsets, a training and a testing subset, with equal number of vectors. A seed is used to generate random mutually exclusive array indices, which are then used to split the data into two ascertaining the training and testing sets are entirely different. The model is trained in an unsupervised fashion to reduce the mean squared error between the input and decoder output and a semi-supervised fashion to learn the the continuous features and class labels. The adversarial networks as well as the autoencoder are trained in three phases: the reconstruction, regularization and semi-supervised phase as mentioned in \cite{makhzani2015adversarial}.  The Adam optimizer \cite{adam_optimizer}, a first-order gradient based optimizer, with a learning rate of 0.001 is used for training in all the phases. In the semi-supervised phase the model is trained to learn the class, position and bandwidth of the input signal by training it on 20\% of the labeled samples from the training set.

\subsection{Implementation details}\label{implementation}
The model is implemented using TensorFlow \cite{tensorflow}, a data flow graph based numerical computation library from Google. Python and C++ bindings of Tensorflow makes the usage of the final trained model easily portable to host based SDR frameworks like GNU Radio \cite{gnuradio_web}. The trained model can be easily imported as a block in GNU Radio which can be readily used in practice with any supported hardware front-end.

\section{Anomaly detection}
\label{anomdetect}
\begin{figure*}[t]
\centering
\input{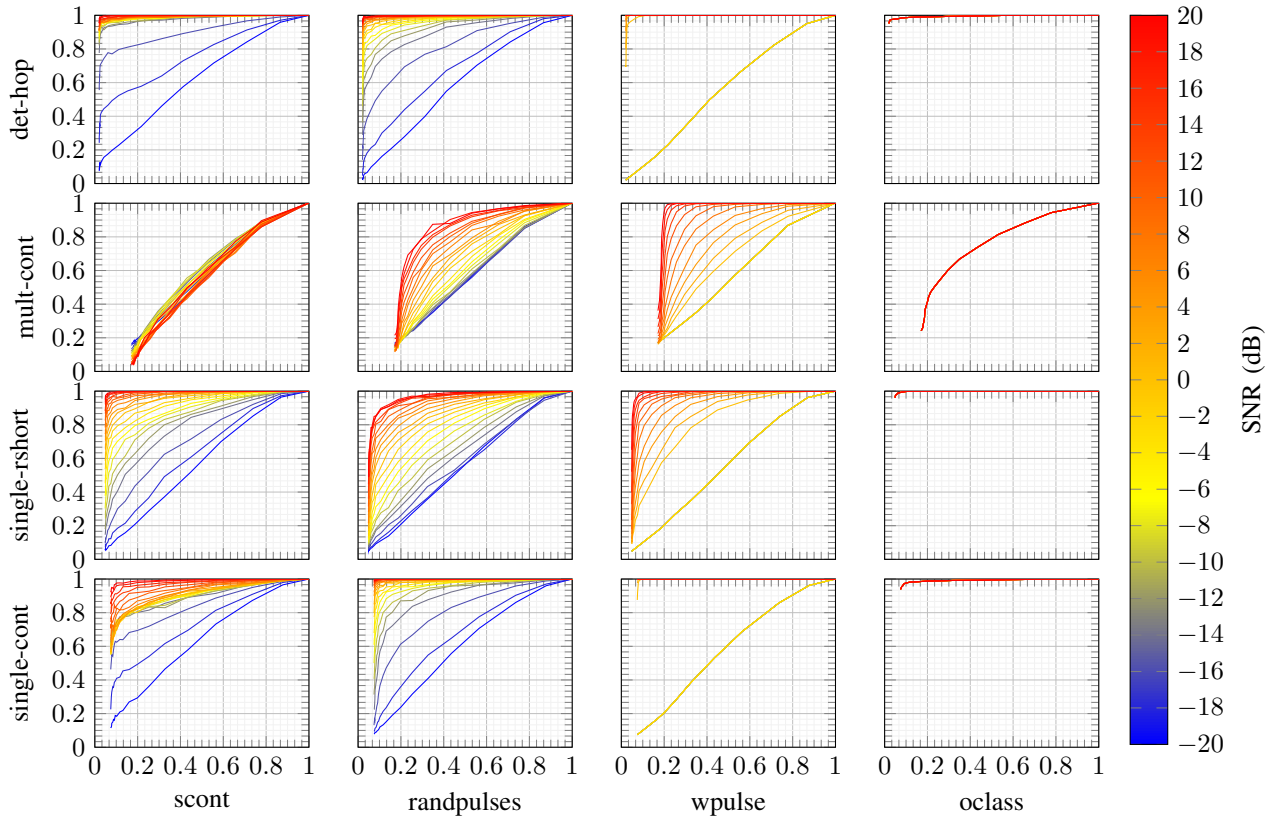}
\caption{ROC curves for different SNRs with synthetic bands in rows and synthetic anomalies in the columns. On each plot the x-axis represents the false positive rate and the y-axis the true positive rate. For \textit{oclass} anomaly, anomaly vectors are randomly selected from other classes and specific SNR based ROC curves are not plotted.}
\label{fig:synthetic_roc}
\end{figure*}

Once the training process is complete, the model weights are frozen and new input data is fed to the model. As mentioned in the model architecture section, anomalies are detected primarily based on the reconstruction error of the model. In addition to the reconstruction error, the classification error and the discriminator loss are also used for detecting anomalous behaviors. 

\subsection{Detection scores}
Three scores are used to detect whether the input data frame is anomalous or not. They are
\begin{enumerate}
\item Reconstruction loss: This error measures the similarity between the input data and the reconstructed data defined as $R_l=\sum\limits_{i=0}^N|\mathbf{x}-\hat{\mathbf{x}}|$ where $\mathbf{x}$ is the frame input, $\hat{\mathbf{x}} = D(\mathbf{z})$ is the decoder frame output and $N$ is the number of data points in the frame.
\item Discriminator loss: The discriminator in the \ac{aae} model is trained to distinguish between the samples from the prior distribution and the samples generated by the encoder. We use the same discrimination loss used during the training process which is defined as $D_l = \sigma(\mathbf{z},1)$ where $\sigma$ is the sigmoid cross entropy. The loss from both continuous ($D_{lcont}$) and categorical ($D_{lcat}$) discriminators are used for computing the final anomaly score.
\item Classification error: The class labels predicted by the encoder is cross checked with the original band of interest for detecting the presence of other known but unexpected signals in a selected frequency band. 
\end{enumerate}
A simple n-sigma threshold is employed on the reconstruction and discriminator loss based on the mean and standard deviation values from the training data. An input data frame is classified as anomalous if $A_{score}$ is $True$:
\begin{multline}
A_{score} = ( R_l > (\mu_{R_{lt}} + n * \sigma_{R_{lt}}))\\
\vee ( (\mu_{D_{ltcont}} - n * \sigma_{D_{ltcont}}) > D_{lcont} > (\mu_{D_{ltcont}} + n * \sigma_{D_{ltcont}}))\\
\vee ( (\mu_{D_{ltcat}} - n * \sigma_{D_{ltcat}}) > D_{lcat} > (\mu_{D_{ltcat}} + n * \sigma_{D_{ltcat}}))\\
\vee ( Class_{Encoder} != Class_{input})
\end{multline}
The threshold value $n$ is selected empirically based on the expected true positive rate and false detection rate. From the probability distributions of \textit{dethop} signal and \textit{dethop} signal with \textit{scont} anomaly shown in Figure~\ref{fig:err_dist}, it can be clearly noticed that the reconstruction loss along with class labels plays a major for anomaly detection.

\begin{figure*}[!t]
\centering
\input{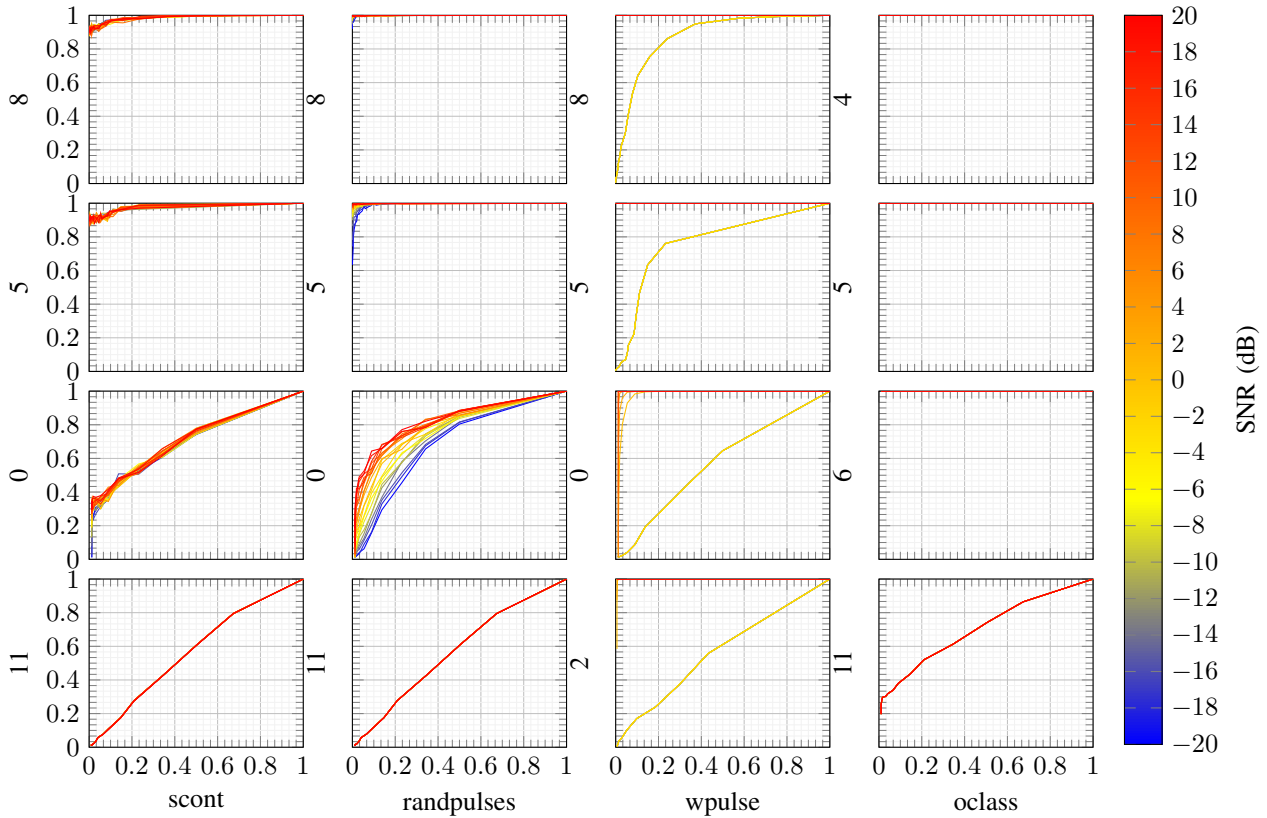}
\caption{ROC curves for the best two (top rows) and worst two (bottom rows) ROC AUC for -20dB SNR in SDR dataset. Synthetic bands with band labels are shown in rows for each synthetic anomaly (columns). On each plot false positive rate is represented on x-axis and true positive rate on y-axis. For \textit{oclass} anomaly, anomaly vectors are randomly selected from other classes and specific SNR based ROC curves are not plotted.}
\label{fig:sdr_roc}
\end{figure*} 

\subsection{Performance comparisons}
To evaluate the performance of the anomaly detector on different datasets, the false alarm rate and the true detection rate are plotted for various frequency bands after injecting different synthetic anomalies. Figure~\ref{fig:synthetic_roc} shows the \ac{roc} curves on the synthetic dataset for different \ac{snr}s. Anomaly signals similar to the original signals are intensionally selected to thoroughly analyze the detection capabilities of the model. For instance, from the \ac{roc} curves it can be seen that detection of \textit{scont} anomaly is difficult in \textit{mult-cont} band as another continuous signal is not an anomalous behavior in the multiple continuous signal band. Similarly detection of \textit{wpulse} works well only on \ac{snr}s above 0~dB as the signal is only visible above the noise floor above 0~dB.

These experiments are repeated on the SDR dataset and the results are plotted in Figure~\ref{fig:sdr_roc}. Only the two best and worst performing frequency bands for different anomalies based on the \ac{auc} for the lowest anomaly \ac{snr} of -20dB are shown due to space limitations, as there are 12 frequency bands in the SDR dataset. Results similar to the synthetic dataset can be noticed in the real capture SDR dataset also. Detecting \textit{scont} and \textit{randpulses} anomalies in frequency band 0 (80-107~MHz) is very difficult as the selected band is very wide and it contains strong FM broadcast stations. Similar results can be noticed in the the other worst performing band 11 (920-960~MHz) which contains GSM signal transmissions that includes both continuous and hopping transmissions. This shows the pressing need to split the 40~MHz bandwidth to multiple bands, for instance continuous and random hopping bands, for better detection accuracies. It can be also noticed that the \textit{oclass} detection accuracies are quite good even in the worst performing band 0 showing the robustness of the signal classification module of the encoder.

\begin{figure}[htb]
\centering
\includegraphics[width=\columnwidth]{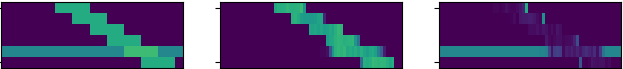}
\includegraphics[width=\columnwidth]{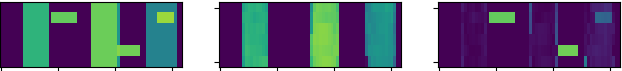}
\includegraphics[width=\columnwidth]{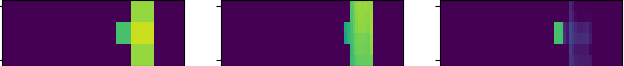}
\caption{Localized anomalies for three different synthetic anomalies. Original input signal, decoder reconstructed signal and the localized anomaly is shown in each row from left to right. First row: \textit{wpulse} anomaly on \textit{dethop} signal, second row: \textit{randpulses} anomaly on \textit{mult-cont} signal, third row: \textit{scont} anomaly on \textit{single-rshort} signal. }
\label{fig:anom_localization}
\end{figure}

\subsection{Anomaly localization}
\begin{figure*}[htb]
\centering
\input{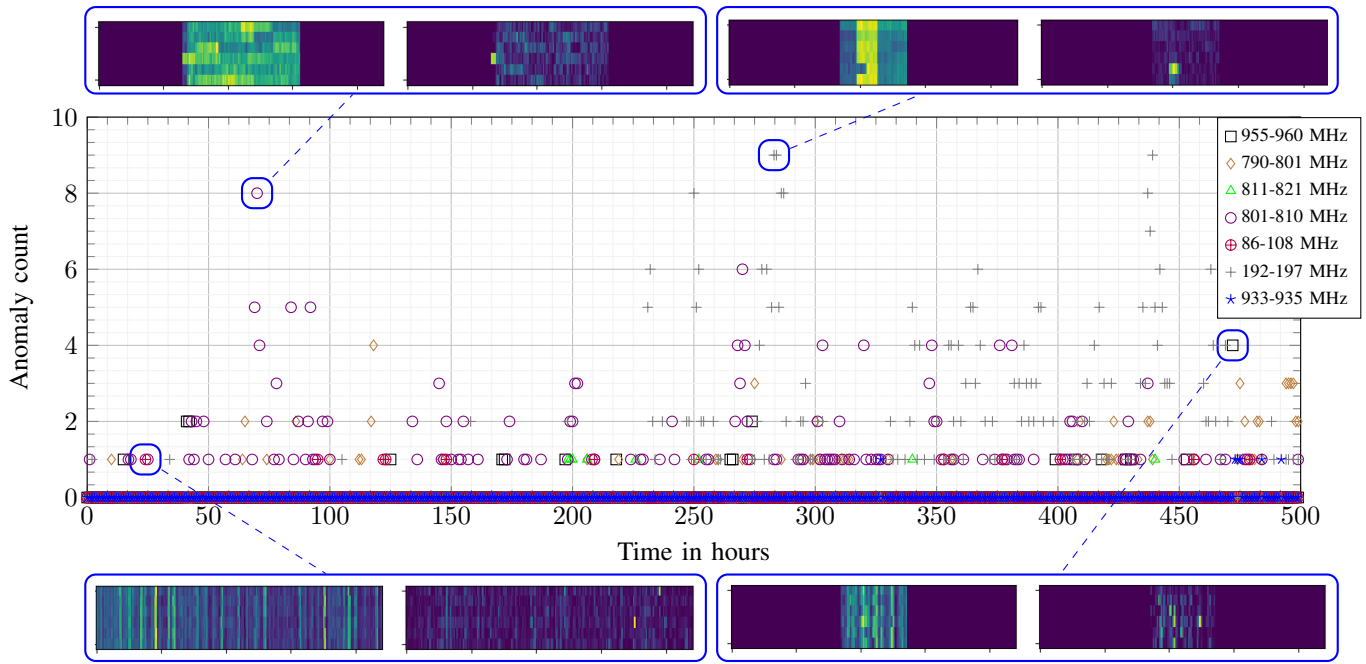}
\caption{Detected anomalies for a duration of 500 hours from one of the Electrosense sensor. Sample input data (left) and the localized anomaly (right) for some sample anomalies are also plotted for some frequency bands.}
\label{fig:electrosense_freerun}
\end{figure*}

\begin{figure}[htb]
\centering
\includegraphics[width=\columnwidth, height=1.2cm]{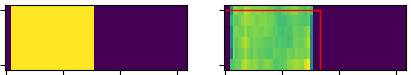}
\includegraphics[width=\columnwidth, height=1.2cm]{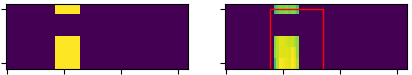}
\includegraphics[width=\columnwidth, height=1.2cm]{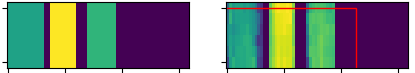}
\caption{Learned bandwidth and position features in a semi-supervised fashion. On each row the left one is the input signal and the right one is the reconstructed signal along with the estimated parameters.} 
\label{fig:semsup_params}
\end{figure}
Localizing anomalies in the wireless frequency spectrum is not common in any of the \ac{soa} algorithms. \ac{saif} presents a simple and robust way to localize the anomalous region from the input \ac{psd} data which is a significant contribution of this paper. In addition to detection of anomalies, the reconstruction error along with the semi-supervised features can be used to localize and understand the anomaly better as shown in Figure~\ref{fig:anom_localization}. Anomaly localization is achieved by plotting the absolute reconstruction error, that is $|\hat{\mathbf{x}}-\mathbf{x}|$. This method works well unless there is a drastic change in the estimated class label which can be noticed in the third row where an \textit{scont} anomaly occurs in an \textit{srshort} band. The model accurately detects it as an anomaly since there is a variation in the estimated class label, but shows the \textit{srshort} signal as the anomalous region instead of \textit{scont}. Figure~\ref{fig:semsup_params} gives some sample plots of the estimated signal position and bandwidth. The current model is only trained for three semi-supervised features including the class label and these interpretable features can be used for analyzing the anomalies better.

\subsection{Anomaly detection in the wild}
To understand the performance on detecting real anomalies, the model is tested on the real-world Electrosense dataset. The model is trained on 7 days of data from one of the Electrosense sensors and tested on the next 500 hours for anomalies with a detection threshold of $3\sigma$~($n=3$). The number of detected anomalies, based on $A_{score}$, along with a few sample anomalies for 7 frequency bands are shown in Figure~\ref{fig:electrosense_freerun}. The model detects unexpected missing transmissions (top-right and bottom-right), high power transmissions (bottom-left) and some out of band transmissions (top-left). It can be noticed that after 230 hours the 192-197~MHz bands started giving more anomalous detections. Visual inspection of the anomalous \ac{psd} patches in this band revealed transmission pattern variations. These detected variations could be either because of the transmitter behavior changes or from the position/antenna changes of the sensor. The model also provides the flexibility to add these anomalous detections to the training set, enabling incremental learning, if the user believes that the behavior is normal. Incorporating this user feedback and enabling automated retraining of models on these kind of anomalous behaviors will be addressed in future work.

\section{Signal compression and classification}
\label{compressclassif}
\begin{figure*}[t]
\centering
\input{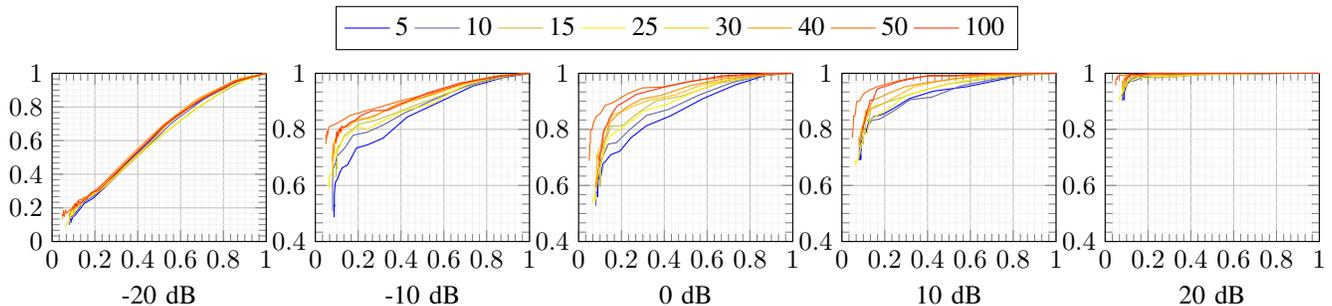}
\caption{\ac{roc} curves for \textit{single-rshort} signal with \textit{scont} anomalies at different \ac{snr}s with varying number of continuous features.}
\label{fig:dimz_var}
\end{figure*} 

To control the data transfer costs associated with the sensing, Electrosense sensors enable three pipelines with very low, medium and high data transfer costs namely: Feature, \ac{psd} and \ac{iq} pipeline. While the \ac{iq} pipeline allows to send raw data to the backend, which can be used to support a broad range of applications, the data transfer rate required is in the 30~Mbps to 100~Mbps range based on the sampling rate of the sensor. The \ac{psd} pipeline on the other hand brings down this rate to hundreds of Kbps. In this section we analyze the compression and classification capabilities of \ac{saif} to reduce the associated data transfer costs.

\subsection{Traditional Spectrum Representation}
In spite of the popularity of various lossy and lossless compression algorithms in image and video processing communities, there are only a few compression algorithms fine tuned for wireless spectrum data. In \cite{hawkins2003new} the authors presented a compression algorithm based on Chebyshev polynomials. The authors in \cite{speccompress} presented a method to separate spectrum noise and other relevant signals specific to L-band satellite signals and then did separate compression to achieve better results when compared to JPEG standards. The aforementioned methods are very specific and lack the compression flexibility when the input data is in multiple formats such as \ac{psd} or \ac{iq}.

\subsection{Non-linear data compression}
Recently unsupervised deep learning models have shown great improvements in compressing input information. In \cite{o2016unsupervised} the authors have achieved 4X to 16X compression ratio on raw sampled \ac{iq} data using autoencoder models. In \ac{saif}, 20 compressed features are used for representing the input \ac{psd} frame. This helps to achieve a lossy compression of 19X, 60X and 120X on the Synthetic, Electrosense and SDR datasets respectively, which can considerably reduce the data transfer costs. Mean absolute reconstruction error of \ac{saif} with 20 features are summarized in Table~\ref{table:classif_recons}. In addition to spectrum reconstruction these features can be used for anomaly detection and signal classification which makes it more attractive. The models can be easily adapted for different data inputs, for instance PSD data in time and frequency or IQ data supporting flexible compression architectures for different sensor data pipelines. The dimension of the compressed feature space along with the model complexity can be adapted to suit the reconstruction loss requirements. For instance, the number of features required to represent the time-frequency \ac{psd} patches of static wireless channels like commercial FM bands will be very less when compared to very random hopping channels. 

An initial analysis is performed, on the synthetic dataset, to understand the trade-off between level of compression and anomaly detection performance by varying the number of continuous features, thereby the compression ratio of the model, which is presented in Figure~\ref{fig:dimz_var}.  At very low (-20~dB) and high (20~dB) anomaly \ac{snr}s there are not much performance gains by increasing the number of features as expected. Detecting signals at -20~dB \ac{snr} is very difficult even with a large number of features whereas at 20~dB \ac{snr} with a smaller set of features can easily detect anomalies due to large variations in the reconstruction loss. While at common \ac{snr} values (-10dB, 0 and 10dB) the anomaly detection performance increases with increasing number of features. We would like to emphasize that the number of features required to achieve reasonable detection performance will depend on the input data dimensions, the encoder and decoder capacity and the dataset complexity itself.

\begin{table}[htb]
\begin{center}
\begin{tabular}{|l|l|l|}
	\hline
    Dataset     & Classification  & Mean absolute\\
                & accuracy (\%) &reconstruction error\\							     
    \hline
  	Synthetic dataset & 92.86 & 7.67 (for 6x64 samples)\\
    \hline
    SDR dataset & 100& 84.12 (for 6x400 samples)\\
    \hline
    Electrosense dataset & 100 & 101.16 (for 6x221 samples)\\
  	\hline
 \end{tabular}
\end{center}
\caption{Band classification accuracy and reconstruction errors on the test data of different datasets.}
\label{table:classif_recons}
\end{table}

\begin{figure}[htb]
\centering
\includegraphics[width=\columnwidth]{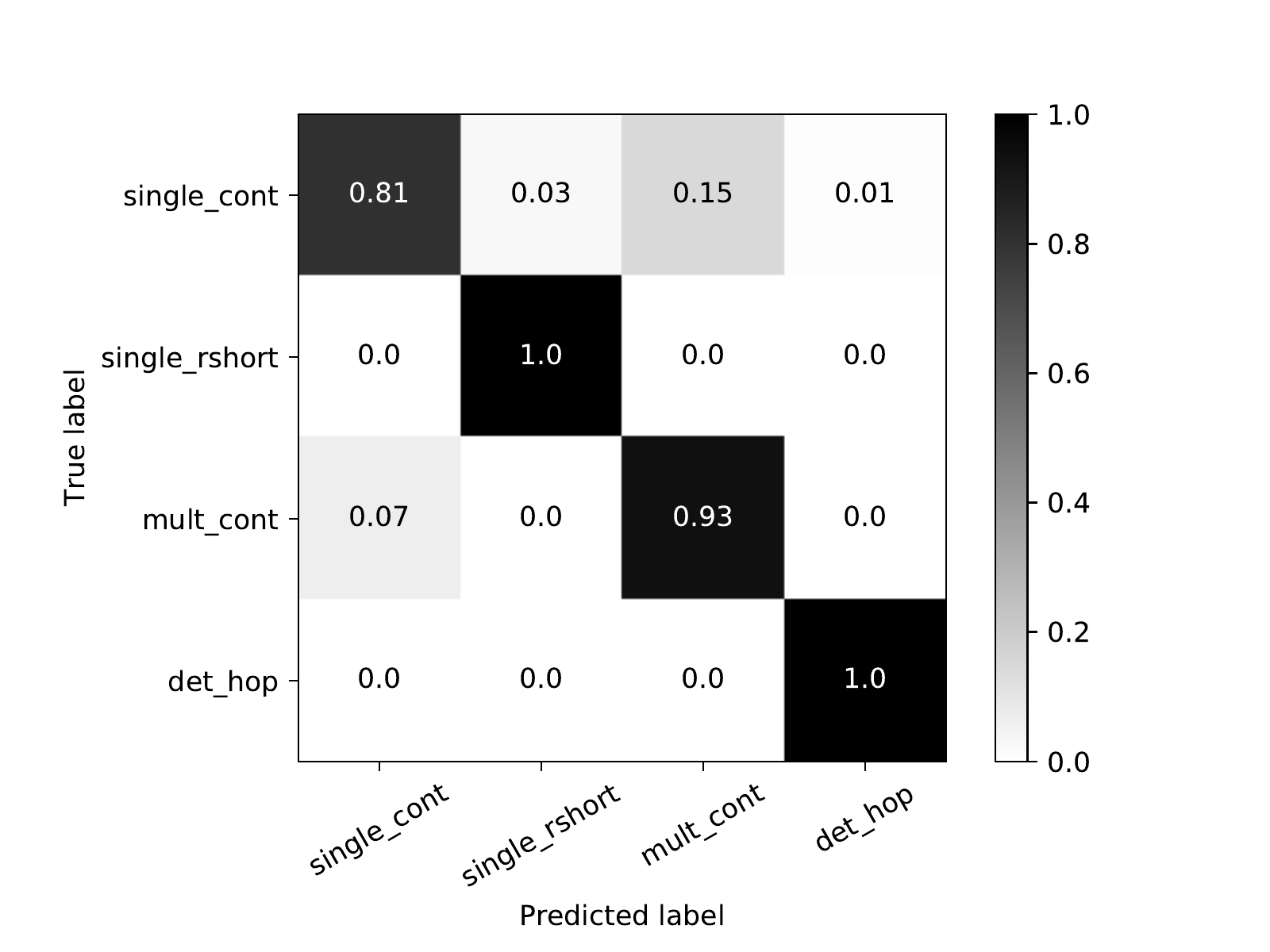}
\caption{Confusion matrix for the synthetic dataset}
\label{fig:confmat_synthetic}
\end{figure}

\subsection{Wireless signal classification}
In addition to anomaly detection \ac{roc} curves, wireless band classification accuracies on the test data of three datasets are summarized on Table~\ref{table:classif_recons}. Since the real wireless bands use different parameters such as signal bandwidths, modulation type, and temporal occupancies at mostly high \ac{snr}s, the wireless band classification problem is not very tough as the classical modulation classification problem \cite{lstm_classif}. On the synthetic dataset the model confuses between \textit{single-cont} and \textit{mult-cont} signal resulting in a classification accuracy of 92.86\%. The confusion matrix for the same is also shown in Figure~\ref{fig:confmat_synthetic}. The model achieves excellent classification accuracy of 100\% on the real SDR and Electrosense dataset. The high classification accuracy stress the fact that a categorical variable helps the encoding process which in-turn helps the decoder to generate fine variations which are specific to a particular class.

\section{Conclusion and Future work}
\label{conclusion}
Automated monitoring of wireless spectrum over frequency, time and space is still a difficult research problem. In this paper we have analyzed the use of an \ac{aae} model in wireless spectrum data anomaly detection, compression and signal classification. We have shown that the proposed model can achieve good anomaly detection and localization along with interpretable feature extraction. The model also can achieve a wireless band classification accuracy close to 100\% by only using 20\% labeled samples. 

In future we would like to perform detailed comparisons of the proposed model with similar prediction based models and also evaluate the performance gains by using raw \ac{iq} samples. Even though we have validated the model performance on one of the Electrosense sensors, we would like to propose some concrete similarity scores that can be used to select closely located or similar spectrum scanning sensors, to enable deployment of a single model across sensors. Further we would like to include user feedback in the entire anomaly detection loop and make the training process fully automated to fulfill the automated spectrum monitoring dream.





%
\balance
\bibliographystyle{IEEEtran}
\bibliography{sections/bibliography}

\end{document}